\documentclass{aastex62}

\usepackage{threeparttable}
\usepackage{longtable}

\shorttitle{Identification and abundance of Galactic HII regions from LAMOST}
\shortauthors{Li-Li Wang et al.}

\begin{document}

\title{Spectroscopic Identification and Chemical Distribution of HII Regions in the Galactic Anti-center Area from LAMOST}

\correspondingauthor{A-Li Luo}
\email{lal@nao.cas.cn}

\author[0000-0002-0786-7307]{Li-Li Wang}
\affil{Key Laboratory of Optical Astronomy, National Astronomical Observatories, Chinese Academy of Sciences, Beijing 100012, China}
\affil{School of Information Management, Dezhou University, Dezhou 253023, China}
\affil{University of Chinese Academy of Sciences, Beijing 100049, China}

\author{A-Li Luo}
\affiliation{Key Laboratory of Optical Astronomy, National Astronomical Observatories, Chinese Academy of Sciences, Beijing 100012, China}

\author{Wen Hou}
\affiliation{Key Laboratory of Optical Astronomy, National Astronomical Observatories, Chinese Academy of Sciences, Beijing 100012, China}

\author{Meng-Xin Wang}
\affil{Key Laboratory of Optical Astronomy, National Astronomical Observatories, Chinese Academy of Sciences, Beijing 100012, China}
\affil{University of Chinese Academy of Sciences, Beijing 100049, China}

\author{Bing Du}
\affil{Key Laboratory of Optical Astronomy, National Astronomical Observatories, Chinese Academy of Sciences, Beijing 100012, China}
\affil{University of Chinese Academy of Sciences, Beijing 100049, China}

\author{Li Qin}
\affil{Key Laboratory of Optical Astronomy, National Astronomical Observatories, Chinese Academy of Sciences, Beijing 100012, China}
\affil{School of Information Management, Dezhou University, Dezhou 253023, China}
\affil{University of Chinese Academy of Sciences, Beijing 100049, China}

\author{Jin-Shu Han}
\affil{School of Information Management, Dezhou University, Dezhou 253023, China}

\begin{abstract}

We spectroscopically identify 101 Galactic HII regions using spectra from the Large Sky Area Multi-Object Fiber Spectroscopic Telescope (LAMOST) survey, cross-matched with an HII region catalog derived from the all-sky \emph{Wide-Field Infrared Survey Explorer}(\emph{WISE}) data. Among all HII regions in our sample, 47 sources are newly confirmed. Spatially, most of our identified HII regions are located in the anti-center area of the Galaxy. For each of the HII regions, we accurately extract and measure the nebular emission lines of the spectra, and estimate the oxygen abundances using the strong-line method. We focus on the abundance distribution of HII regions in the Galactic anti-center area. Accordingly, we derive the oxygen abundance gradient with a slope of -0.036$\pm$0.004 dex $\rm kpc^{-1}$, covering a range of $R_G$ from 8.1 to 19.3 kpc. In particular, we also fit the outer disk objects with a slope of -0.039$\pm$ 0.012 dex $\rm kpc^{-1}$, which indicates that there is no flattening of the radial oxygen gradient in the outer Galactic disk.
\end{abstract}


\section{Introduction} \label{sec:intro}
HII regions are ionized gaseous nebula formed by ultraviolet radiation from massive stars. The spectra of HII regions are characterized by rich emission lines of hydrogen and metal elements. The chemical abundances of Galactic HII regions represent the gas-phase abundances of the Milky Way today. Especially, oxygen is a proxy for the abundance analysis of ionized gaseous nebula, which is usually derived from the emission lines at optical bands. The abundance distribution across the Galactic disk is crucial to place constraints on theories of the formation and evolution of the Milky Way \citep{Searle1971,Shaver1983}.

Although there are many observations and studies of HII regions in the Milky Way, the sampling of Galactic HII regions remains incomplete. In particular, HII regions in the Galactic anti-center used for the studies of abundance gradient are relatively few \citep{FernandezMart2017}. Recently, \cite{Anderson2014} provided a relatively complete Galaxy HII region catalog based on the \emph{Wide-Field Infrared Survey Explorer} (\emph{WISE}) project, and the catalog has been updated by \cite{Anderson2015,Anderson2018}. This catalog contains $\sim$8,400 entries for two types of objects: known HII regions and candidate HII regions, all of which share the bubble morphology in \emph{WISE} images. In this catalog, $\sim$1,500 sources are located in the anti-center of the Milky Way. However, approximately 73\% are candidate HII regions that cannot be confirmed as true HII regions. \cite{Anderson2014} claimed that some of these candidates are ideal objects for spectroscopic observations.

The spectroscopic survey of the Galactic anti-center is an important part of the Large Sky Area Multi-Object Fiber Spectroscopic Telescope (LAMOST) survey, which provides abundant data to study the Galactic HII regions in the direction of the anti-center. In this study, we cross-match LAMOST optical data with the \emph{WISE} catalog of Galactic HII Regions to identify the candidate HII regions from a spectroscopic perspective, aimed at increasing the number of spectra of HII regions toward the Galactic anti-center.

Using optical, infrared, or radio data, different authors revealed the negative O/H abundance gradients from -0.04 to -0.07 dex $\rm kpc^{-1}$ for HII regions at Galactocentric distances of ${R_G}=$0--22 kpc \citep{Hawley1978,Shaver1983,Afferbach1995,Deharveng2000,Esteban2005,Rudolph2006,Balser2015,FernandezMart2017,Esteban2017}. In particular, the radial abundance gradients of the outer Galaxy have been emphasized. Some studies observed a flatter radial gradient at the outer part of the Milky Way \citep{Fich1991,Vilchez1996,Korotin2014}, whereas others claimed the absence of such flattening \citep{Deharveng2000,Rudolph2006}. Most recently, \cite{FernandezMart2017} presented the chemical compositions of 23 Galactic anti-center HII regions located at ${R_G}>$11kpc and did not confirm the flattening of the distribution in the outer disk. \cite{Esteban2017} performed deep optical spectroscopy of eight HII regions located in the anti-center and studied the radial oxygen gradients defined by the HII regions within ${R_G}$  of 11.5--17 kpc; their results indicated the absence of flattening in the radial oxygen gradient at the outer part of the Galactic disk.

This study has two main purposes. The first is to identify the HII regions in the Galactic anti-center area using LAMOST spectra in order to expand the current spectra samples of HII regions. The second is to study the chemical abundance gradient of HII regions in the Galactic anti-center area and the outer disk using our optical data.

The outline of this article is as follows. We present the data and sample selection in Section 2. Section 3 describes the process of identifying HII regions using LAMOST spectra. The main results of our analysis are included in Section 4, where we describe the spectroscopic properties of the HII regions and oxygen abundance gradients in the Galactic anti-center area and the outer disk. Finally, our main conclusions are summarized in Section 5.

\section{Data and sample selection}

We select our sample based on LAMOST spectroscopic data and the \emph{WISE} HII region catalog.

\subsection{Data in LAMOST DR5}

The LAMOST survey is a spectroscopic survey that covers the northern sky. The telescope is a 4-meter reflecting Schmidt telescope with a $5^{\circ}$ field of view and 4000 fibers are almost evenly distributed over the focal plane. The diameter of a fiber is 3 arcsec. In addition, the telescope has 16 spectrographs and 32 charge coupled device (CCD) cameras (each spectrograph is equipped with two CCD cameras of a blue arm from 3,700\AA \ to 5,900\AA \ and a red arm from 5,700\AA \ to 9,000\AA), and hence, there are 250 fiber spectra in each obtained CCD image \citep{Cui2012,Zhao2012,Luo2015}. Its spectral resolution is R$\sim$1800. In June 2017, the first five-year mission of LAMOST was completed. On December 31, 2017, the fifth data release (DR5) of LAMOST was officially released for domestic astronomers and international collaborators. The Galactic spectroscopic survey is the main part of LAMOST Survey, which observes celestial objects over the entire available northern sky. As a unique component of LAMOST Galactic survey, the Galactic anti-center survey covers Galactic longitudes $150^{\circ}$ $\leq$ $\ell$ $\leq$ $210^{\circ}$ and latitudes $|b|\leq$ $30^{\circ}$ \citep{Yuan2015}. There are millions of spectra located in this area in LAMOST DR5, which provide abundant data for the identification of HII regions in the Galactic anti-center direction.

Our study utilizes the sky spectra in the LAMOST dataset to detect Galactic HII regions. According to the requirements of LAMOST observations, approximately 20 fibers in each spectrograph are assigned to point toward blank sky, where the sky spectra are extracted for data reduction. If the telescope points toward an HII region, the nebular emission features of the HII region may be directly superposed upon the sky spectra. Therefore, the sky spectra offer immense potential for detecting HII regions.

\subsection{\emph{WISE} HII region catalog}\label{sec:WISE_HII}

The \emph{WISE} HII region catalog was compiled by \cite{Anderson2014}, and updated by \cite{Anderson2015,Anderson2018} as a part of ongoing HII Region Discovery Survey\citep{Bania2010,Anderson2011}. We use the most recent version of the \emph{WISE} HII region catalog: \emph{WISE} Catalog V2.0(hereafter HIICat\_V2) from http://astro.phys.wvu.edu/wise. The catalog includes more than 8,400 known Galactic HII regions and candidate HII regions by searching for their characteristic mid-infrared (MIR) morphology, using data from the \emph{WISE} satellite. The catalog extends over all Galactic longitudes within $|b|\leq$ $8^{\circ}$ and five high-mass star-forming regions at high Galactic latitudes. There are approximately 1,900 sources (labeled as ``K'') in the catalog defined as known HII regions as they have been observed in RRL or H$\alpha$ spectroscopic emission, whereas $\sim$6,500 sources are HII region candidates. Among these candidates, $\sim$2,800 sources (labeled as ``C'' and ``G'') have radio continuum and MIR emission, and are ideal objects for follow-up spectroscopic observations to confirm their classification, whereas $\sim$3,700 sources (labeled as ``Q'') have the characteristic MIR emission of HII regions, but do not show radio continuum emission at the sensitivity limits of the existing surveys. For the HII region candidates, further observations are required to confirm whether they are true HII regions. In this study, we use the spectra from the LAMOST dataset to confirm some candidates as HII regions, especially toward the Galactic anti-center.

The angular radii of HII regions in HIICat\_V2 range from 6$^{\prime\prime}$ to 1$^\circ$6, which approximately contain the MIR emission of each source. The average angular radii of known and candidate HII regions are both 100$^{\prime\prime}$ \citep{Anderson2014}. In addition, the HIICat\_V2 provides Galactocentric distances for $\sim$ 2,700 sources. Most of the distances are determined by kinematic method, only approximately 7\% of sources use parallax distances. The detailed distance determination was presented in \cite{Anderson2014}. They computed kinematic distances using the Brand(1986) rotation curve model, with the Sun 8.5 kpc from the Galactic center and a solar circular rotation speed of 220 km s$^{-1}$. Sources in the inner Galaxy suffer from the kinematic distance ambiguity (KDA), which complicates the computation of kinematic distances. A KDA resolution(KDAR) requires auxiliary data to determine. They employed three methods for resolving the KDA: HI Emission Absorption (HI E/A), H$_{2}$CO absorption, and HI self-absorption (HI SA). They also provided the average distance uncertainties: $\sim$15\% in the first and fourth quadrants and $\sim$30\% in the second and third Galactic quadrants.

\subsection{Data reduction of LAMOST spectra}

We match the LAMOST spectroscopic catalog to the positions of the HIICat\_V2, and obtain $\sim$3,000 LAMOST spectra with good quality, which are located in the areas of known HII regions and candidate HII regions presented in the HIICat\_V2. Using the standard data reduction via two-dimensional (2D) pipeline of the LAMOST survey, including fiber tracing, flux extraction, wavelength calibration, and flat fielding (further details in \citealt{Luo2015}), the spectra in our sample are extracted as one-dimensional (1D) data.

After the above reduction, background subtraction is performed. In the 2D pipeline of LAMOST, the sky background is modeled for each spectrograph and thereafter subtracted from the spectra in the same spectrograph. If the spectrograph points toward an HII region, the nebular emissions that may be superimposed on the sky background are thereafter subtracted from the spectra. Hence, to find the ``clean'' sky background for the spectra falling in a HII region candidate, we select sky spectra outside but adjoined to the range of this HII region candidate. The clean sky spectra have very weak or even no nebular emission features such as [NII]$\lambda$$\lambda$6549,6585 and [SII]$\lambda$$\lambda$6718,6732. The sky backgrounds are then subtracted from the spectra in our sample.

\begin{figure}
\centering
\includegraphics[width=12cm]{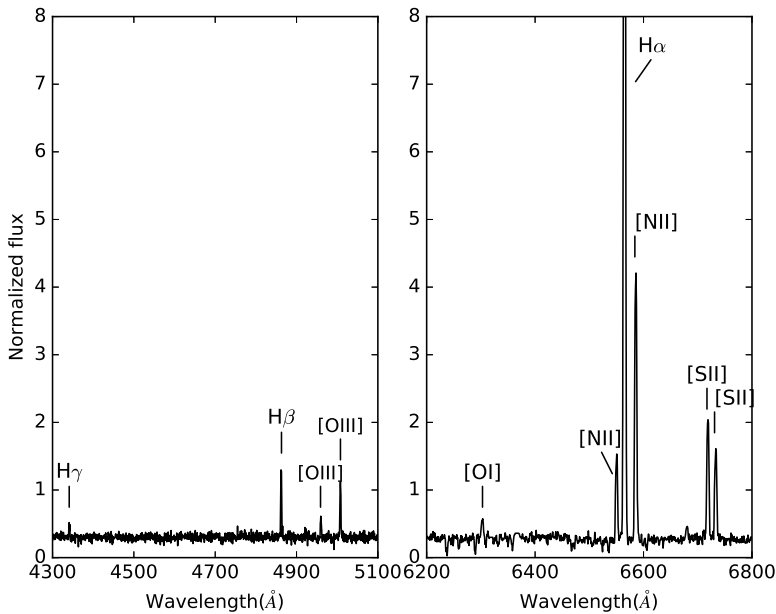}
\caption{Sections of a 1D spectrum obtained from LAMOST after data reduction in our sample. The prominent nebular lines are labeled as markers: H$\gamma$, H$\beta$, [OIII]$\lambda$4960,[OIII]$\lambda$5007, [OI]$\lambda$6302, H$\alpha$, [NII]$\lambda$$\lambda$6549,6585, and [SII]$\lambda$$\lambda$ 6718,6732. We have not corrected for the interstellar extinction.}
\label{fig:example_spec}
\end{figure}

We do not perform an absolute flux calibration because the LAMOST project is designed as a spectroscopic survey without photometric standard stars. Fortunately, we mainly focus on the ratios of emission lines used to identify HII regions and compute the chemical abundance. As an example, Figure \ref{fig:example_spec} shows an extracted 1D spectrum in our sample. Notable major emission features are marked at their rest wavelengths.

\subsection{Sample selection}

In order to search for the spectra with emission features of HII regions, we detect whether the spectra have [SII]$\lambda$$\lambda$6718,6732 emissions, which are not only important nebular features, but are also uncontaminated by sky background and starlight. We present a subjective criterion for detecting [SII]$\lambda$$\lambda$6718,6732 in emissions. The criterion is shown in Equation \ref{eq:eq1}.

\begin{equation}
 {\rm max} ( flux [\lambda_{1} - 1 , \lambda_{1} + 1]) \ge \mu_1 +2*\sigma_1
 \ \ {\rm and}\ \  {\rm max} ( flux[ \lambda_{2} - 1,\lambda_{2} + 1]) \ge \mu_2 + 2*\sigma_2
\label{eq:eq1}
\end{equation}

where $flux$ represents the spectral flux pixel in the wavelength range around the [SII] double lines, $\lambda_{1}$ and $\lambda_{2}$ are the center wavelengths of [SII]($\lambda_{1}=6718 \AA$, $\lambda_{2}=6732  \AA$), max($flux$) represents the maximum flux around the line center $\pm$1\AA, and $\mu$ and $\sigma$ are the mean and standard deviation of the fluxes in the wavelength range around the line center, respectively, and the wavelength range we used is line center $\pm$4\AA(namely, $\lambda_{1}$$\pm$4\AA \ and $\lambda_{2}$$\pm$4\AA). Using Equation \ref{eq:eq1} we can identify [SII]$\lambda$$\lambda$6718,6732 in emissions if the fluxes of line centers are larger than 2$\sigma$ of the fluxes in their neighborhoods. We further visually inspect the spectra one-by-one, abandoning the spectra if the main spectral lines (H$\alpha$, [NII]$\lambda$6585, and [SII]$\lambda$$\lambda$6718,6732) are severely affected by noise.

According to this criterion and visual inspection, $\sim$1,000 LAMOST spectra with [SII]$\lambda$$\lambda$6718,6732 emissions are selected. In some candidate HII regions, up to a few dozen spectra are located in the same HII region. In this case, we extract an averaged spectrum for this region. We use the median method for averaging, because the median spectrum can preserve the relative fluxes of emission features without altering the emission line ratios \citep{Vanden Berk2001}, which is very important for HII region identification and chemical abundance calculation. Thus, in the current study we focus on the understanding of the average properties of the HII regions using our spectra.

\section{Identification of HII regions}

One of our goals is to spectroscopically confirm HII regions. In this section, we describe the procedures employed to identify HII regions using the spectra in our sample. First, we measure the emission lines of the spectra. Subsequently, we identify the HII regions using an emission-line diagnostic diagram based on log([SII]$\lambda$$\lambda$6718,6732/H$\alpha$) versus log([NII]$\lambda$6585/H$\alpha$)\citep{Kniazev2008}.

\subsection{Emission-line measurements}

We calculate the relative intensities of lines(H$\beta$, [OIII]$\lambda$5007, H$\alpha$, [NII]$\lambda$$\lambda$6549,6585, and [SII]$\lambda$$\lambda$6718,6732) by integrating all the fluxes in the wavelength range. H$\alpha$ and [NII]$\lambda$$\lambda$6549,6585 are deblended with two or three Gaussian profiles. In order to measure the lines more accurately, the center and width of the Gaussian profile are adjusted simultaneously and automatically in the fitting process to avoid the deviation caused by redshift measurement. This automatic adjustment has been used effectively for emission-line measurements in galaxy spectra in our previous study \citep{Wang L.L.2018}.

We catalog the flux ratios of the emission lines in Table \ref{tab:catalogue}. [NII]$\lambda$6585/H$\alpha$ and [SII]$\lambda$$\lambda$ 6718,6732)/H$\alpha$ are presented in this table and used for the following HII region identification and abundance calculation. Note that approximately 50\% of our spectra have weak or undetected H$\beta$ and/or [OIII]$\lambda$5007, in which case we do not provide the flux ratios of these two lines.

\subsection{Emission-line diagnostic of HII regions}

There are several classes of ionized nebulae, such as planetary nebulae (PNe), HII regions, and supernova remnants, whose spectra have similar emission lines but different line intensity ratios. Empirical emission-line diagnostic diagrams are often used to classify the ionized nebulae in the Galaxy \citep{Sabbadin1977,RiesgoTirado2002,Magrini2003,Kniazev2008,Lagrois2012}. \cite{Kniazev2008} analyzed a series of classification diagrams and used two criteria to distinguish PNe from HII regions. These two criteria (Equation (1) and (3) in their paper) are based on log([OIII]$\lambda$5007/H$\beta$) versus log([NII]$\lambda$6585/H$\alpha$) and log([SII]$\lambda$$\lambda$ 6718,6732/ H$\alpha$) versus log([NII]$\lambda$6585/H$\alpha$). To perform the identification of HII regions, we employ their quantitative criterion based on log([SII]$\lambda$$\lambda$ 6718,6732/ H$\alpha$) versus log([NII]$\lambda$6585/H$\alpha$) to separate HII regions from PNe. And also, this criterion is useful for rejecting possible supernovae remnants(SNRs). Note that the PN and SNR classifications are outside the scope of the current study, we use this diagnostic diagram in order to separate HII regions from other ionized nebulae. We do not employ another criterion as some of our spectra do not have H$\beta$ and/or [OIII]$\lambda$5007. The classification criterion we use is as follows:

\begin{equation}
 \log \left( {\frac{{\left[ {{\rm{SII}}} \right]}}{{{\rm{H\alpha }}}}} \right) \ge 0.63\log \left( {\frac{{\left[ {{\rm{NII}}} \right]}}{{{\rm{H\alpha }}}}} \right) - 0.55
\label{eq:eq2}
\end{equation}

\begin{figure}
\centering
\includegraphics[width=10cm]{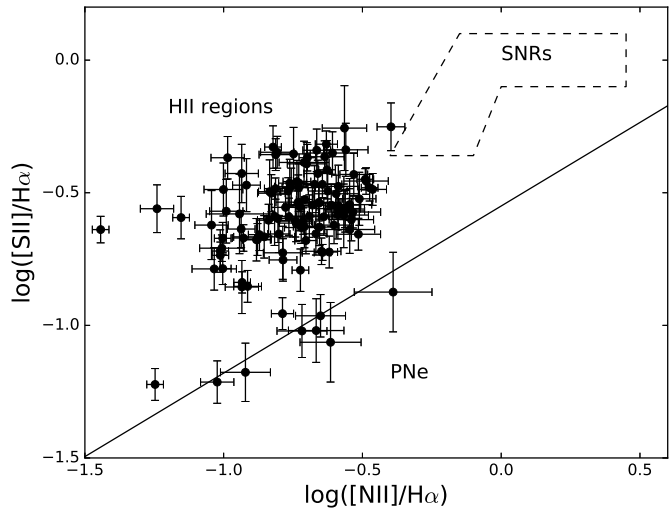}
\caption{ Classification diagram for the ionized nebulae: HII regions, PNe, and SNRs. The solid line graphically shows the criterion for HII/PNe separation that is used in this work. The loci of SNRs is shown with the dashed line on the right top of the figure.}
\label{fig:diagram_class}
\end{figure}

Using Equation \ref{eq:eq2}, we distinguish 101 HII regions from other types of nebulae. The diagnostic diagram is shown in Figure \ref{fig:diagram_class}. The loci of HII regions, PNe and SNRs are separated by different lines. There are several points located in the PNe area. We leave these points out of account because these points are very close to the border line ,also with high uncertainties of line ratios.

Among all the identified HII regions in our sample, 47 sources are newly confirmed as HII regions, defined as candidate HII regions in the HIICat\_V2(labeled as ``C,'' ``G,'' and ``Q''), whereas others are known HII regions (labeled as ``K''). In Figure \ref{fig:Plot_cover}, the red and cyan circles represent the Galactic locations of the newly confirmed and previously known sources, respectively. The newly confirmed HII regions expand the current spectra samples of HII regions, and the other sources in our sample spectroscopically verify the previously known ones using radio or infra data.

\section{Analysis and results}

In this section, we compile the spectroscopic properties of the identified HII regions into a catalog, and explore the oxygen abundance and radial gradient of HII regions in the Galactic anti-center area and the outer disk.

\subsection{Catalog of the HII regions}

\begin{figure*}
\includegraphics[width=18cm]{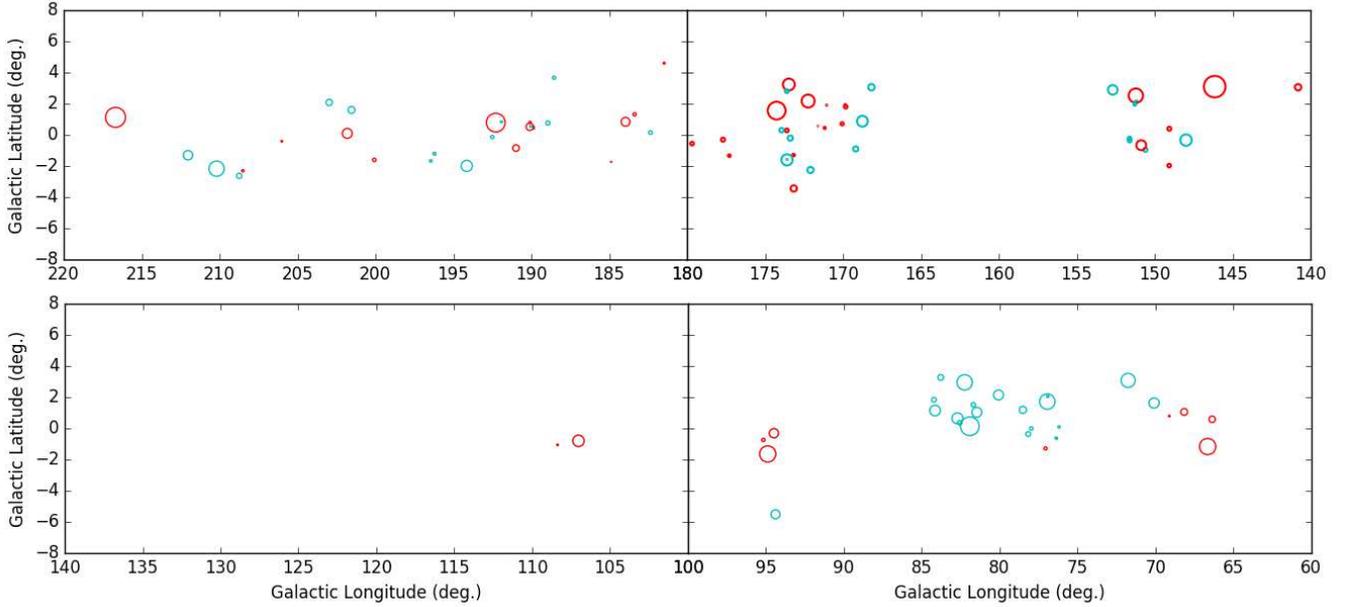}
\caption{Spatial distribution of the HII regions identified in this study. The red and cyan circles show the locations of the newly confirmed and previously known sources, respectively.}
\label{fig:Plot_cover}
\end{figure*}

The Galactic locations of the HII regions identified in this study are shown in Figure \ref{fig:Plot_cover}. We can see that this sample is mainly in the region of the Galactic anti-center. Approximately 76.2\% of all sources are located in the second and third Galactic quadrants (41.6\% and 34.6\%, respectively), and 23.8\% are located in the first quadrant. No source is located in the fourth quadrant, which is due to the observational strategy of the LAMOST survey.

We create a catalog for the spectroscopic parameters of the identified HII regions in our sample, which is shown in Table \ref{tab:catalogue}. The first six columns are obtained from HIICat\_V2: the name of the HII region, Galactic longitude, Galactic latitude, approximate circular radius in degrees, $R_{G}$ which indicates the Galactocentric radius in kpc, and the classification of HII regions by Anderson et al. which is detailed in Section~\ref{sec:WISE_HII}. We compute the equivalent width of H$\alpha$ emission line ($EW$\_H$\alpha$) for HII regions. We also provide three typical line ratios: [OIII]$\lambda$5007/H$\beta$, [NII]$\lambda$6585/H$\alpha$, and [SII]$\lambda$$\lambda$ 6718,6732/H$\alpha$, and the last two items are used in the classification diagnostic. Some entities in this table have no [OIII]$\lambda$5007/H$\beta$ if H$\beta$ and/or [OIII]$\lambda$5007 have not been detected in our spectra. The last column in Table \ref{tab:catalogue} indicates the oxygen abundances and their uncertainties derived using strong emission lines, as presented in detail in Section~\ref{sec:OHbylines}. Our newly confirmed HII regions are marked with `*' at the first column of 47 entries in this table.

\begin{longtable}[]{lllllllllll}
\caption{Catalog for the Spectroscopic Properties of HII Regions in Our Sample.}  \label{tab:catalogue} \\
\hline \hline
Name  & $l$   & $b$   &Radius & $R_{G}$ & Cat. & EW\_H$\alpha$  &\multicolumn3c{Line Ratios}  &12+log(O/H) \\
\cline{8-10}
      & (deg) & (deg) & (deg) & (kpc) &  & (\AA)  & [OIII]/H$\beta$  & [NII]/H$\alpha$ & [SII]/H$\alpha$ &  \\
\hline
\endfirsthead
\multicolumn{11}{c}%
{{\bfseries \emph \tablename\ \thetable{}}{ \emph {(continued)}}} \\
\hline \hline
Name  & $l$   & $b$   &Radius & $R_{G}$ & Cat. & EW\_H$\alpha$  &\multicolumn3c{Line Ratios}  &12+log(O/H) \\
\cline{8-10}
      & (deg) & (deg) & (deg) & (kpc) &  & (\AA)  & [OIII]/H$\beta$  & [NII]/H$\alpha$ & [SII]/H$\alpha$ & \\
\hline
\endhead
\hline \multicolumn{11}{c}{{\emph \tablename\ \thetable{}}{\emph { continued on next page}}} \\
\endfoot
\endlastfoot
G066.375+00.583* & 66.375  & 0.583   & 0.20 & --    & Q & -6.23(0.99)  & --         & 0.18(0.05) & 0.35(0.11) & 8.48(0.06) \\
G066.659$-$01.167* & 66.659   & -1.167  & 0.52 & --    & Q & -5.84(1.14)  & --         & 0.17(0.05) & 0.26(0.07) & 8.46(0.08) \\
G068.168+01.058* & 68.168  & 1.058   & 0.22 & --    & G & -9.09(1.66)  & --         & 0.24(0.07) & 0.32(0.11) & 8.54(0.08) \\
G069.118+00.793* & 69.118  & 0.793   & 0.05 & --    & C & -6.83(1.25)  & 0.47(0.15) & 0.28(0.08) & 0.46(0.11) & 8.58(0.07) \\
G070.099+01.629 & 70.099    & 1.629   & 0.33 & 10.3 & K & -21.64(3.90) & 0.32(0.11) & 0.29(0.08) & 0.55(0.16) & 8.60(0.08) \\
G071.764+03.076 & 71.764  & 3.076   & 0.45 & 8.1  & K & -81.51(5.71) & 0.33(0.12) & 0.30(0.07) & 0.27(0.08) & 8.60(0.07) \\
G076.197+00.092 & 76.197  & 0.092   & 0.08 & 8.3  & K & -8.68(1.46)  & --         & 0.25(0.06) & 0.24(0.07) & 8.56(0.08) \\
G076.309$-$00.667* & 76.309   & -0.667  & 0.02 & --    & Q & -8.09(1.34)  & --         & 0.22(0.08) & 0.29(0.09) & 8.53(0.09) \\
G076.384$-$00.621 & 76.384  & -0.621  & 0.08 & 8.3  & K & -11.07(1.70) & 0.45(0.16) & 0.21(0.06) & 0.23(0.06) & 8.51(0.08) \\
G076.917+02.054 & 76.917  & 2.054   & 0.07 & 8.3  & K & -59.87(5.99) & 0.95(0.25) & 0.24(0.06) & 0.29(0.07) & 8.55(0.07) \\
G076.951+01.718 & 76.951  & 1.718   & 0.50 & 8.3  & K & -27.15(5.41) & 0.90(0.27) & 0.35(0.10) & 0.27(0.07) & 8.64(0.08) \\
G077.059$-$01.285* & 77.059   & -1.285  & 0.10 & --    & Q & -11.38(2.16) & 0.77(0.21) & 0.19(0.06) & 0.33(0.09) & 8.48(0.08) \\
G077.977$-$00.004 & 77.977  & -0.004  & 0.11 & 8.3  & K & -10.42(1.72) & --         & 0.27(0.07) & 0.24(0.07) & 8.57(0.07) \\
G078.177$-$00.363 & 78.177  & -0.363  & 0.15 & 8.3  & K & -11.37(2.17) & --         & 0.21(0.07) & 0.23(0.07) & 8.51(0.09) \\
G078.516+01.187 & 78.516  & 1.187   & 0.23 & 8.3  & K & -12.79(2.14) & --         & 0.28(0.08) & 0.31(0.09) & 8.58(0.08) \\
G080.079+02.145 & 80.079  & 2.145   & 0.32 & 8.4  & K & -8.46(1.57)  & --         & 0.33(0.08) & 0.29(0.07) & 8.63(0.07) \\
G081.470+01.030 & 81.470   & 1.030   & 0.32 & 8.4  & K & -16.94(2.64) & 0.32(0.11) & 0.20(0.06) & 0.25(0.06) & 8.49(0.08) \\
G081.920+00.138 & 81.920  & 0.138   & 0.60 & 8.4  & K & -20.38(3.06) & 0.64(0.20) & 0.41(0.11) & 0.35(0.09) & 8.68(0.07) \\
G082.253+02.953 & 82.253  & 2.953   & 0.49 & 8.4  & K & -15.77(2.54) & --         & 0.28(0.08) & 0.26(0.06) & 8.58(0.08) \\
G082.566+00.362 & 82.566  & 0.362   & 0.14 & 8.4  & K & -7.29(1.14)  & --         & 0.24(0.11) & 0.33(0.09) & 8.55(0.11) \\
G082.721+00.646 & 82.721  & 0.646   & 0.35 & 8.4  & K & -32.05(4.17) & 0.42(0.16) & 0.25(0.07) & 0.26(0.08) & 8.56(0.08) \\
G083.792+03.269 & 83.792  & 3.269    & 0.18 & 8.5  & K & -53.33(4.80) & 0.31(0.11) & 0.34(0.10) & 0.33(0.09) & 8.64(0.08) \\
G084.157+01.146 & 84.158  & 1.146   & 0.34 & 13.6 & K & -12.15(1.82) & 0.53(0.19) & 0.16(0.04) & 0.18(0.05) & 8.45(0.08) \\
G084.221+01.835 & 84.221  & 1.835   & 0.15 & 14.4 & K & -10.88(1.66) & 0.33(0.15) & 0.19(0.05) & 0.25(0.06) & 8.49(0.08) \\
G094.492$-$00.310* & 94.493  & -0.310   & 0.30 & --    & C & -10.56(2.09) & 0.17(0.05) & 0.15(0.05) & 0.33(0.09) & 8.44(0.08) \\
G094.890$-$01.643* & 94.890   & -1.642  & 0.52 & --    & G & -25.37(3.30) & 0.55(0.13) & 0.10(0.03) & 0.18(0.05) & 8.32(0.07) \\
G095.168$-$00.739* & 95.169  & -0.739  & 0.11 & --    & C & -10.38(1.67) & --         & 0.20(0.07) & 0.43(0.15) & 8.50(0.08) \\
G107.034$-$00.801 & 107.034 & -0.801  & 0.37 & 10.6 & K & -18.24(3.04) & 0.34(0.12) & 0.15(0.04) & 0.21(0.06) & 8.44(0.08) \\
G108.375$-$01.056 & 108.375 & -1.056  & 0.05 & 11.7 & K & -11.61(1.95) & 0.37(0.13) & 0.09(0.03) & 0.20(0.05) & 8.30(0.08) \\
G140.793+03.052* & 140.793 & 3.053   & 0.21 & --    & C & -10.58(1.87) & --         & 0.04(0.01) & 0.23(0.05) & 8.08(0.08) \\
G146.144+03.093* & 146.144 & 3.093   & 0.69 & --    & C & -12.06(2.23) & --         & 0.06(0.02) & 0.28(0.08) & 8.19(0.07) \\
G147.984$-$00.335 & 147.984 & -0.335  & 0.37 & 10.1 & K & -6.52(1.0)   & --         & 0.25(0.08) & 0.19(0.05) & 8.56(0.08) \\
G149.053+00.393* & 149.053 & 0.393   & 0.13 & --    & Q & -8.93(1.74)  & --         & 0.09(0.02) & 0.24(0.05) & 8.31(0.07) \\
G149.063$-$01.974* & 149.063 & -1.973  & 0.11 & --    & Q & -5.51(0.86)  & --         & 0.19(0.06) & 0.34(0.09) & 8.48(0.08) \\
G150.596$-$00.955 & 150.596 & -0.955  & 0.13 & 11.6 & K & -19.23(3.73) & 1.09(0.30) & 0.12(0.03) & 0.15(0.03) & 8.37(0.08) \\
G150.860$-$00.666* & 150.860  & -0.666  & 0.32 & --    & C & -14.46(2.44) & 0.52(0.19) & 0.23(0.06) & 0.48(0.13) & 8.54(0.07) \\
G151.180+02.124 & 151.180  & 2.124   & 0.08 & 16.8 & K & -9.41(1.54)  & --         & 0.10(0.03) & 0.16(0.05) & 8.32(0.07) \\
G151.205+02.516* & 151.205 & 2.517   & 0.46 & --    & G & -8.87(1.55)  & --         & 0.10(0.03) & 0.16(0.04) & 8.33(0.07) \\
G151.273+01.961 & 151.273 & 1.961   & 0.09 & 12.1 & K & -8.39(1.34)  & --         & 0.15(0.04) & 0.14(0.05) & 8.43(0.07) \\
G151.606$-$00.373 & 151.606 & -0.372  & 0.12 & 19.3 & K & -6.10(1.13)  & --         & 0.04(0.02) & 0.21(0.08) & 8.11(0.10) \\
G151.609$-$00.233 & 151.609 & -0.232  & 0.12 & 17.3 & K & -12.40(2.03) & 2.01(0.69) & 0.07(0.03) & 0.21(0.06) & 8.23(0.09) \\
G152.687+02.889 & 152.687 & 2.889   & 0.31 & 10.4 & K & -8.49(1.39)  & --         & 0.16(0.05) & 0.19(0.06) & 8.44(0.08) \\
G168.171+03.054 & 168.171 & 3.055   & 0.21 & --    & K & -10.74(1.93) & 0.62(0.19) & 0.13(0.04) & 0.22(0.05) & 8.40(0.07) \\
G168.750+00.873 & 168.750  & 0.874   & 0.35 & --    & K & -13.15(1.58) & 0.50(0.15)  & 0.16(0.05) & 0.22(0.06) & 8.44(0.07) \\
G169.180$-$00.905 & 169.180  & -0.904  & 0.17 & 13.8 & K & -6.60(0.95)  & --         & 0.13(0.03) & 0.34(0.08) & 8.39(0.07) \\
G169.823+01.803* & 169.823 & 1.803   & 0.11 & --    & C & -7.88(1.24)  & --         & 0.15(0.04) & 0.26(0.06) & 8.43(0.07) \\
G169.851+01.930* & 169.851 & 1.930    & 0.05 & --    & C & -8.91(1.55)  & --         & 0.16(0.04) & 0.25(0.07) & 8.44(0.07) \\
G170.051+00.708* & 170.051 & 0.709   & 0.10 & --    & Q & -5.21(0.76)  & --         & 0.07(0.02) & 0.25(0.07) & 8.24(0.06) \\
G171.046+01.906* & 171.046 & 1.906   & 0.03 & --    & Q & -6.14(0.77)  & --         & 0.16(0.04) & 0.45(0.11) & 8.44(0.07) \\
G171.171+00.445* & 171.171 & 0.446   & 0.07 & --    & Q & -10.32(1.66) & --         & 0.17(0.04) & 0.32(0.09) & 8.46(0.06) \\
G171.602+00.563* & 171.602 & 0.563   & 0.02 & --    & Q & -5.28(1.0)   & --         & 0.12(0.03) & 0.34(0.08) & 8.38(0.06) \\
G172.080$-$02.258 & 172.080  & -2.257  & 0.20 & --    & K & -55.65(5.01) & 0.17(0.05) & 0.28(0.06) & 0.23(0.06) & 8.59(0.06) \\
G172.231+02.167* & 172.231 & 2.168   & 0.42 & --    & C & -6.98(1.17)  & --         & 0.12(0.03) & 0.23(0.05) & 8.37(0.06) \\
G173.156$-$03.442* & 173.156 & -3.442  & 0.20 & --    & Q & -8.01(1.42)  & --         & 0.18(0.05) & 0.33(0.10) & 8.47(0.07) \\
G173.168$-$01.299* & 173.168 & -1.299  & 0.08 & --    & C & -16.44(3.04) & 0.29(0.09) & 0.32(0.09) & 0.36(0.09) & 8.62(0.07) \\
G173.375$-$00.209 & 173.375 & -0.208  & 0.17 & --    & K & -18.36(2.81) & 0.16(0.05) & 0.17(0.05) & 0.28(0.07) & 8.46(0.07) \\
G173.468+03.230* & 173.468 & 3.230    & 0.38 & --    & G & -22.36(2.91) & --         & 0.19(0.06) & 0.28(0.07) & 8.49(0.07) \\
G173.582$-$01.581 & 173.582 & -1.58   & 0.04 & --    & K & -65.70(5.91) & 1.55(0.40) & 0.12(0.03) & 0.14(0.04) & 8.37(0.06) \\
G173.588$-$01.606 & 173.588 & -1.606  & 0.36 & --    & K & -43.42(3.91) & 0.98(0.25) & 0.16(0.05) & 0.19(0.05) & 8.45(0.07) \\
G173.598+00.281* & 173.598 & 0.282   & 0.12 & --    & Q & -8.65(1.60)  & --         & 0.17(0.06) & 0.34(0.11) & 8.47(0.08) \\
G173.599+02.803 & 173.599 & 2.804   & 0.12 & --    & K & -25.91(3.11) & --         & 0.24(0.06) & 0.39(0.10) & 8.54(0.06) \\
G173.937+00.298 & 173.937 & 0.298   & 0.14 & --    & K & -17.45(2.74) & --         & 0.29(0.07) & 0.37(0.10) & 8.60(0.06) \\
G174.254+01.555* & 174.254 & 1.555   & 0.57 & --    & C & -10.54(1.84) & --         & 0.15(0.04) & 0.32(0.09) & 8.42(0.07) \\
G177.291$-$01.339* & 177.291 & -1.338  & 0.09 & --    & C & -8.23(1.21)  & --         & 0.10(0.03) & 0.43(0.11) & 8.34(0.07) \\
G177.696$-$00.311* & 177.696 & -0.311   & 0.13 & --    & C & -5.96(0.74)  & --         & 0.20(0.05) & 0.26(0.08) & 8.51(0.06) \\
G179.682$-$00.564* & 179.682 & -0.564  & 0.12 & --    & Q & -6.42(1.03)  & --         & 0.15(0.04) & 0.47(0.12) & 8.43(0.07) \\
G181.471+04.599* & 181.471 & 4.599     & 0.06 & --    & Q & -6.26(0.92)  & --         & 0.21(0.06) & 0.28(0.08) & 8.51(0.07) \\
G182.349+00.148 & 182.349 & 0.148   & 0.12 & --    & K & -40.32(4.44) & --         & 0.28(0.07) & 0.28(0.08) & 8.59(0.06) \\
G183.370+01.316* & 183.370  & 1.316   & 0.10 & --    & Q & -7.27(1.12)  & --         & 0.12(0.03) & 0.37(0.08) & 8.37(0.07) \\
G183.947+00.835* & 183.947 & 0.835   & 0.28 & --    & Q & -9.41(1.59)  & --         & 0.15(0.04) & 0.31(0.08) & 8.43(0.06) \\
G184.870$-$01.733* & 184.870  & -1.733  & 0.03 & --    & Q & -7.79(1.21)  & --         & 0.15(0.04) & 0.44(0.10) & 8.44(0.06) \\
G188.539+03.666 & 188.539 & 3.667   & 0.10 & --    & K & -8.39(1.38)  & 0.50(0.18) & 0.22(0.06) & 0.37(0.11) & 8.52(0.07) \\
G188.935+00.755 & 188.935 & 0.756   & 0.13 & --    & K & -18.84(2.31) & --         & 0.20(0.05) & 0.30(0.07) & 8.50(0.06) \\
G189.830+00.417 & 189.830  & 0.418   & 0.06 & 10.6 & K & -199.93(12.0)& 0.21(0.09) & 0.20(0.06) & 0.29(0.07) & 8.51(0.08) \\
G189.859+00.502* & 189.859 & 0.502   & 0.05 & --    & G & -40.06(4.81) & 0.87(0.22) & 0.23(0.05) & 0.19(0.05) & 8.53(0.06) \\
G190.049+00.538 & 190.049 & 0.539   & 0.05 & --    & K & -72.83(3.64) & 1.82(0.54) & 0.06(0.02) & 0.06(0.02) & 8.19(0.07) \\
G190.075+00.795* & 190.075 & 0.795   & 0.07 & --    & G & -34.49(3.79) & 0.65(0.18) & 0.31(0.08) & 0.30(0.08) & 8.61(0.06) \\
G190.095+00.516* & 190.095 & 0.517   & 0.23 & --    & G & -74.87(5.24) & 0.76(0.19) & 0.24(0.07) & 0.19(0.05) & 8.55(0.07) \\
G190.977$-$00.851* & 190.977 & -0.851   & 0.21 & --    & Q & -6.53(1.02)  & --         & 0.11(0.03) & 0.26(0.06) & 8.36(0.07) \\
G191.916+00.837 & 191.916 & 0.837   & 0.06 & --    & K & -11.44(1.79) & --         & 0.23(0.06) & 0.43(0.10) & 8.54(0.06) \\
G192.274+00.783* & 192.274 & 0.784   & 0.60 & --    & Q & -8.30(1.66)  & --         & 0.10(0.03) & 0.27(0.09) & 8.34(0.07) \\
G192.499$-$00.142 & 192.499 & -0.141  & 0.11 & --    & K & -38.94(5.06) & --         & 0.43(0.12) & 0.56(0.15) & 8.69(0.07) \\
G194.138$-$01.996 & 194.138 & -1.995  & 0.36 & --    & K & -29.15(3.79) & 0.17(0.05) & 0.27(0.07) & 0.26(0.06) & 8.58(0.07) \\
G196.163$-$01.255* & 196.163 & -1.254  & 0.03 & --    & Q & -6.10(0.88)  & 0.60(0.21) & 0.22(0.06) & 0.46(0.12) & 8.52(0.06) \\
G196.220$-$01.204 & 196.220  & -1.203  & 0.09 & --    & K & -16.57(2.79) & 0.21(0.06) & 0.25(0.06) & 0.45(0.09) & 8.55(0.06) \\
G196.448$-$01.673 & 196.448 & -1.673  & 0.08 & 13.7 & K & -8.48(1.30)  & 0.45(0.14) & 0.15(0.04) & 0.44(0.12) & 8.42(0.08) \\
G200.070$-$01.609* & 200.070  & -1.609  & 0.11 & --    & C & -7.26(1.13)  & 1.81(0.44) & 0.19(0.05) & 0.41(0.08) & 8.50(0.06) \\
G201.535+01.597 & 201.535 & 1.597   & 0.22 & 12.2 & K & -20.04(2.53) & 1.34(0.43) & 0.29(0.09) & 0.28(0.08) & 8.59(0.09) \\
G201.806+00.091* & 201.806 & 0.091   & 0.32 & --    & Q & -6.82(1.09)  & 0.93(0.30) & 0.23(0.06) & 0.34(0.10) & 8.53(0.07) \\
G202.968+02.083 & 202.968 & 2.083   & 0.20 & 9.2  & K & -9.83(1.56)  & 0.55(0.14) & 0.30(0.09) & 0.27(0.08) & 8.60(0.08) \\
G206.008$-$00.413* & 206.008 & -0.412  & 0.05 & --    & Q & -12.30(2.41) & 0.87(0.29) & 0.20(0.06) & 0.41(0.13) & 8.50(0.07) \\
G206.466$-$16.349 & 206.466 & -16.349 & 0.28 & 9.1  & K & -7.03(1.28)  & --         & 0.31(0.07) & 0.22(0.05) & 8.61(0.07) \\
G207.068$-$16.256* & 207.068 & -16.256 & 0.05 & --    & Q & -22.99(4.20) & --         & 0.26(0.07) & 0.28(0.10) & 8.57(0.07) \\
G208.445$-$18.994* & 208.445 & -18.994 & 0.30 & --    & Q & -16.51(2.97) & --         & 0.09(0.02) & 0.16(0.05) & 8.31(0.07) \\
G208.506$-$02.304* & 208.506 & -2.304  & 0.07 & --    & Q & -7.18(1.37)  & --         & 0.26(0.07) & 0.28(0.08) & 8.57(0.07) \\
G208.741$-$02.634 & 208.741 & -2.633  & 0.17 & 9.7  & K & -22.48(3.70) & 0.71(0.21) & 0.27(0.08) & 0.22(0.06) & 8.57(0.08) \\
G209.037$-$19.377 & 209.037 & -19.376 & 0.15 & 8.9  & K & -185.99(11.16)& 1.79(0.48)& 0.16(0.05) & 0.11(0.03) & 8.45(0.08) \\
G209.107$-$19.510 & 209.107 & -19.509 & 0.48 & 8.9  & K & -19.61(3.27) & 0.67(0.18) & 0.29(0.08) & 0.25(0.08) & 8.59(0.08) \\
G210.187$-$02.169 & 210.187 & -2.168  & 0.49 & 10.7 & K & -9.83(1.83)  & --         & 0.25(0.08) & 0.24(0.08) & 8.56(0.09) \\
G212.021$-$01.309 & 212.021 & -1.309  & 0.30 & 14.4 & K & -7.97(1.20)  & 0.72(0.19) & 0.15(0.04) & 0.25(0.08) & 8.42(0.08) \\
G216.673+01.122* & 216.673 & 1.123   & 0.64 & --    & Q & -9.93(1.39)  & --         & 0.13(0.04) & 0.21(0.06) & 8.40(0.08) \\
\noalign{\smallskip}\hline
\end{longtable}
\tablecomments{The first six columns are obtained from HIICat\_V2. The sixth column is the classification of HII regions presented by Anderson et al., detailed in Section~\ref{sec:WISE_HII}. The emission line [OIII] represents [OIII]$\lambda$5007, [NII] represents [NII]$\lambda$6585, and [SII] represents [SII]$\lambda$$\lambda$ 6718, 6732. If H$\beta$ or [OIII]$\lambda$5007 have not been detected in our spectra, the line ratios [OIII]/H$\beta$ are labeled as ``-''. The lines in table have not been corrected for the interstellar extinction. The records marked with `*' at the first column are our newly confirmed HII regions.}

\subsection{Oxygen abundance}\label{sec:OHbylines}

Chemical abundance is an important property that can be derived from the emission lines in the optical wavelength range. The easiest element to measure in the HII region emission spectra is oxygen \citep{Lopez-Sanchez2012}. The O/H abundance in HII regions is typically expressed in terms 12+log(O/H). Two typical methods are used to determine oxygen abundance: the direct method and strong-line method. The direct method requires the measurement of the electron temperature ($T_e$) based on the auroral lines such as [OIII]$\lambda$4363, [NII]$\lambda$5755, and [SII]$\lambda$6312\citep{Garnett1992,Stasinska2005,Esteban2017,FernandezMart2017}. However, these auroral lines are very weak and cannot be observed in galaxies or HII regions without very sensitive, high-S/N spectra\citep{Kewley2008}. Therefore, strong-line indicators, such as $R_{23}$\citep{Pagel1979}, N2O2\citep{Kewley2002}, N2\citep{van Zee1998,PP2004,Marino2013}, and O3N2\citep{Alloin1979,PP2004,Marino2013} were developed as tracers of oxygen abundance, through empirical calibrations from $T_e$ metallicities and theoretical calibrations using photoionization models. These calibrations were summarized by \cite{Kewley2008}. \cite{PerezMontero2017} presented a detailed review of the determination of chemical abundances in gaseous nebulae using the direct and strong-line methods.

To evaluate the oxygen abundance of our HII regions, we adopt the strong-line method rather than the $T_e$ method as the auroral [OIII]$\lambda$4363 are too faint to detect in our spectra. $R_{23}$ and N2O2 line ratios are not used in our work, because they are both based on [OII]$\lambda$3727, which is noisy in our spectra owing to the smaller throughput at the blue end of the spectra. For 50\% of our objects, H$\beta$ or [OIII]$\lambda$5007 are very weak or undetected, and therefore, the fraction of HII regions with O3N2 ratios is only half of the total number in our sample. Therefore, we employ N2 as the estimator, which is expressed as Equation \ref{eq:eqN2}, for the purpose of the current study. N2 is more insensitive to uncertainties of extinction correction and flux calibration, because it is measured from the close proximity of lines only in red spectra rather than from separately obtained blue and red spectra. Generally, N2 is valid for the range -2.5 $<$ N2 $<$ -0.3 \citep{PP2004}. The values of this indicator in our sample are within this range.

\begin{equation}
N2=log([NII]\lambda6585/H\alpha)
\label{eq:eqN2}
\end{equation}

To estimate the oxygen abundance using the indicator N2, we adopt the widely used empirical calibration relation proposed by \cite{PP2004}:

\begin{equation}
12 + log(O/H) = 8.90 + 0.57 * N2
\label{eq:eqOH}
\end{equation}
where N2 is defined as Equation \ref{eq:eqN2}. In general, this calibration is valid for -2.5 $<$ N2 $<$ -0.3. Hereafter, we refer to this relation as PP04\_N2.

The values of oxygen abundance of HII regions in our sample are between 8.08 and 8.70, which are displayed in Table \ref{tab:catalogue}. For 50\% of objects in our sample, the oxygen abundances are greater than 8.5, and there are no very low abundances (12+log(O/H)$<$8.0), which indicates that the HII regions in our sample are of moderate metallicities.

\begin{figure}
\centering
\includegraphics[width=10cm]{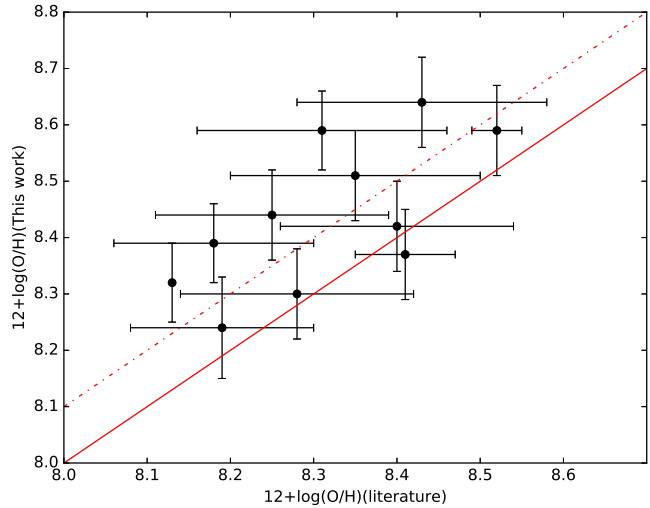}
\caption{ Comparison between the oxygen abundances based on the $T_e$ method and our abundance values for the same sources in our sample. The red solid line is the equal-value line, and the dashed line is the linear least-square fit to the data points, which shows a systematic offset of 0.1 dex and a scatter of 0.15 dex. }
\label{fig:compare_OH}
\end{figure}

Some studies claimed that the oxygen abundance derived by Equation \ref{eq:eqOH} based on the N2 indicator has a high uncertainty, owing to the lack of a parameter that considers the ionization degree of the gas \citep{PP2004, Lopez-Sanchez2012}. The $T_e$ method is believed to be the more accurate method for abundance determinations. Hence, we compare the $T_e$-based oxygen abundances with our results for the same sources in our sample to test the validity of our abundance determinations for HII regions. Table \ref{tab:compare_OH} lists 11 Galactic HII regions by cross-matching objects in our sample with the ones in the literature whose oxygen abundances have been derived using the $T_e$ method. The comparison of 12+(O/H) is graphically shown in Figure \ref{fig:compare_OH}. We can see that our oxygen abundance values are systematically $\sim$ 0.1 dex higher than the values determined by the $T_e$ method, and the scatter is $\sim$ 0.15 dex. Similar scatters were also found by \cite{Yin2007} and \cite{Lopez-Sanchez2012}. The systematic offset is consistent with the previous studies that claimed that the $T_e$-based oxygen abundances may underestimate
the abundance when temperature fluctuations exist in the emission-line nebulae \citep{Stasinska2005, Bresolin2007,Lopez-Sanchez2012}.

\begin{table}[h!]
\renewcommand{\thetable}{\arabic{table}}
\centering
\caption{Comparison of 12+(O/H) between Our Abundance Values and $T_e$-based Ones from the Literature} \label{tab:compare_OH}
\begin{tabular}{ccccc}
\tablewidth{0pt}
\hline
\hline
Name & Other Name & 12+(O/H)(This Work) & 12+(O/H)(Literature) & Reference \\
\hline
G070.099+01.629 & S100 & 8.60 & 8.52 & 1 \\
G071.764+03.076 & S101 & 8.60 & 8.31 & 2 \\
G083.792+03.269 & S112 & 8.64 & 8.43 & 2 \\
G107.034$-$00.801 & S142 & 8.44 & 8.25 & 2 \\
G108.375$-$01.056 & S148 & 8.30 & 8.28 & 2 \\
G150.596$-$00.955 & S206 & 8.37 & 8.44 & 3 \\
G151.180+02.124 & S207 & 8.32 & 8.13 & 3 \\
G151.609$-$00.233 & S209 & 8.23 & 8.19 & 3 \\
G169.180$-$00.905 & S228 & 8.39 & 8.18 & 3 \\
G189.830+00.417 & S252 & 8.51 & 8.35 & 2 \\
G212.021$-$01.309 & S284 & 8.42 & 8.40 & 2 \\
\hline
\end{tabular}
\begin{tablenotes}
\item[1]Notes: The first column is obtained from HIICat\_V2. The second column is the other name in the reference. The numbers in the last column represent different values from the literature: 1--\cite{Esteban2017}; 2--\cite{Rudolph2006}; 3--\cite{FernandezMart2017}.
\end{tablenotes}
\end{table}

\subsection{Radial gradient at the anti-center}\label{sec:gradients}

Another aim of this study is to investigate the radial gradient of oxygen abundance in the Galactic anti-center direction. In this section, we explore the gradient of the abundance distribution using HII regions in our sample.

\begin{figure}
\centering
\includegraphics[width=12cm]{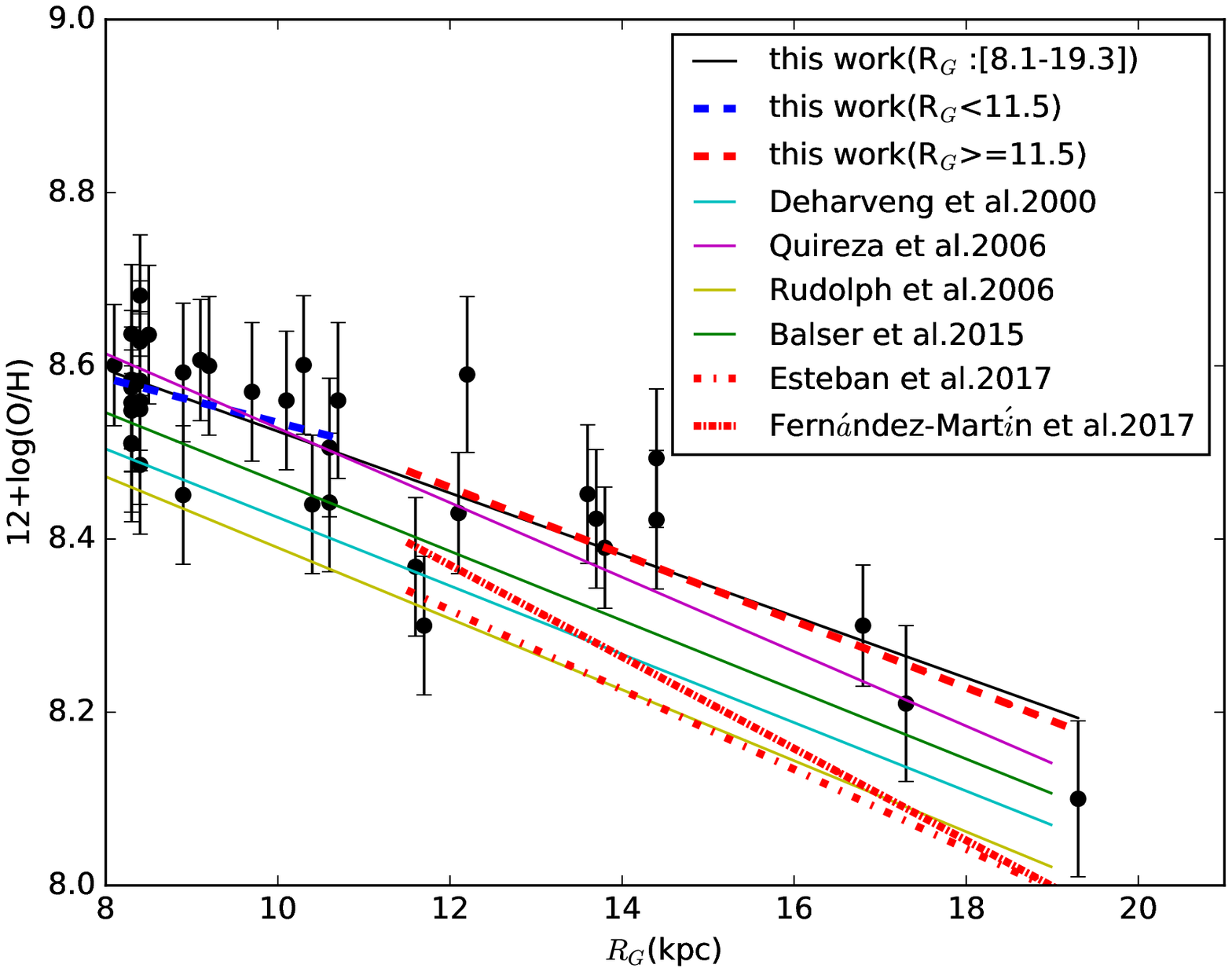}
\caption{Radial gradients of 12+(O/H) versus Galactocentric distance $R_G$(kpc) in the range 8.1 to 19.3 kpc. The solid black line represents the least-squares fit to all objects in our sample with a gradient of -0.036($\pm$0.004)dex $\rm kpc^{-1}$. The blue and red dashed line are the fits to the HII regions in our sample with $R_G<$11.5 kpc and $R_G$ $\geq$11.5 kpc, respectively. The other solid lines with different colors and the two dash-dotted red lines represent the extrapolated gradients derived from the literature.}
\label{fig:metal_gradients}
\end{figure}

In our sample, we obtain approximately 38 objects with Galactocentric distance in the range of 8.1--19.3 kpc from the HIICat\_V2. There are 12 objects lying in the outer part of the Milky Way ($R_G$ $\geq$11.5 kpc). Figure \ref{fig:metal_gradients} shows the chemical abundance distribution of O/H in our sample. We perform a least-squares linear fit of the oxygen abundance versus the Galactocentric radius $R_G$ in the range 8.1--19.3, which is plotted as a black solid line in the figure. The radial oxygen abundance gradient is expressed as

\begin{equation}
12 + {\rm log(O/H)} = 8.88(\pm0.04)-0.036(\pm0.004)R_G
\label{eq:OH_grad_all}
\end{equation}

When comparing our gradient of -0.036$\pm$0.004 dex $\rm kpc^{-1}$ with those obtained from the literature for a similar range of distances which are shown in solid lines with different colors in Figure \ref{fig:metal_gradients}, we find that our gradient is consistent with the gradients obtained by \cite{Deharveng2000}(-0.0395$\pm$0.0049 dex $\rm kpc^{-1}$) and \cite{Balser2015}(-0.04 dex $\rm kpc^{-1}$). The gradients obtained by \citep{Quireza2006}(-0.043$\pm$0.007dex $\rm kpc^{-1}$) and \cite{Rudolph2006}(-0.041$\pm$0.014 dex $\rm kpc^{-1}$) are steeper than our result, consistent with our determination within errors.

We also perform two additional least-squares linear fits to the HII regions with $R_G<$11.5 kpc and $R_G$ $\geq$11.5 kpc in order to compare the abundance distributions in the inner and outer parts of the Milky Way, which are plotted as blue and red dash lines in Figure \ref{fig:metal_gradients}, respectively. The radial gradients for HII regions in the inner and outer Galaxy are expressed as Equation \ref{eq:OH_grad_in} and Equation \ref{eq:OH_grad_out}, respectively.

\begin{equation}
12 + {\rm log(O/H)} = 8.79(\pm0.12)-0.027(\pm0.013)R_{G,<11.5}
\label{eq:OH_grad_in}
\end{equation}

\begin{equation}
12 + {\rm log(O/H)} = 8.92(\pm0.17)-0.039(\pm0.012)R_{G,\geq11.5}
\label{eq:OH_grad_out}
\end{equation}

The most recent determinations of oxygen gradient in the Galactic anti-center are presented in \cite{FernandezMart2017} and \cite{Esteban2017}, which are graphically demonstrated in dash-dotted red lines in Figure \ref{fig:metal_gradients}. \cite{FernandezMart2017} obtained a gradient of -0.053$\pm$0.009 dex $\rm kpc^{-1}$ using 23 HII regions with the distance $R_G$ from 11 to 18 kpc, and \cite{Esteban2017} derived a gradient of -0.046$\pm$0.017 dex $\rm kpc^{-1}$ with $R_G$ from 11.5 to 17.0 kpc. We obtained a gradient of -0.039$\pm$0.012 dex $\rm kpc^{-1}$ in a similar range of distances($R_G$:[11.5-19.3] kpc), which is shallower than their gradients but is within the uncertainty.

The abundance comparison between the inner and outer parts of the Galaxy supports the Inside-Out formation scheme of the disk \citep{Chiappini1997,Korotin2014}. The metal abundances in the inner part are higher than those in the outer part of disk because the inner disk is usually formed first and the outer disk is formed progressively later with a relatively long time scale for star formation and metal enrichment. The gradient variations can be used to determine whether the flattening of the gradient exists at the outer disk. Some works indicate that the abundance gradient of the outer part is flatter than that of the inner part \citep{Fich1991,Vilchez1996,Korotin2014}, whereas others confirm the absence of flattening in the radial oxygen abundance gradient at the outer disk \citep{Deharveng2000,Rudolph2006,FernandezMart2017,Esteban2017}. In our work, our data suggest that there is no flattening of the oxygen abundance gradient in the outer disk of the Galaxy, at least up to 19.3 kpc.

\section{Summary and Conclusions}

The objective of this work was to extend the optical spectra sample of HII regions at the Galactic anti-center and study the chemical abundance gradient based on these data. We presented the spectroscopic identifications of 101 HII regions based on the \emph{WISE} HII region catalog from http://astro.phys.wvu.edu/wise, among which 47 sources were newly confirmed. We first selected the spectra with [SII]$\lambda$$\lambda$ 6718,6732 emission lines, and thereafter classified the HII regions from other types of gaseous nebulae using an emission-line diagnostic based on [SII]$\lambda$$\lambda$ 6718,6732/ H$\alpha$ versus [NII]$\lambda$6585/H$\alpha$. Spatially, most of our HII regions were located in the anti-center direction of the Milky Way, which provides opportunities for studying the radial gradient of chemical abundance in the Galactic anti-center area.

We determined the oxygen abundances for all objects in our sample using a strong-line indicator, i.e., N2. Among all the HII regions in our sample, the Galactocentric distances of 38 objects were obtained from the \emph{WISE} HII region catalog, covering a range of $R_G$ from 8.1 to 19.3 kpc. Using these objects, we derived a least-square linear fit to the oxygen abundance gradient with -0.036$\pm$0.004 dex $\rm kpc^{-1}$. we also fitted the outer disk objects with a gradient of -0.039$\pm$0.012 dex $\rm kpc^{-1}$. This result demonstrates the absence of flattening of the radial gradient of oxygen abundance in the outer part of the Milky Way.




\acknowledgments

This work is supported by the National Key Basic Research Program of China (Grant No. 2014CB845700), the National Natural Science Foundation of China (Grant Nos. 11390371 and 11673005), and the Joint Research Fund in Astronomy (Grant No. U1531119) under cooperative agreement between the National Natural Science Foundation of China (NSFC) and the Chinese Academy of Sciences (CAS).

The Guo Shou Jing Telescope (the Large Sky Area Multi-Object Fiber Spectroscopic Telescope, LAMOST) is a National Major Scientific Project built by the Chinese Academy of Sciences. Funding for the project has been provided by the National Development and Reform Commission. LAMOST is operated and managed by National Astronomical Observatories, Chinese Academy of Sciences.




\begin{thebibliography}{}
\expandafter\ifx\csname natexlab\endcsname\relax\def\natexlab#1{#1}\fi
\providecommand{\url}[1]{\href{#1}{#1}}
\providecommand{\dodoi}[1]{doi:~\href{http://doi.org/#1}{\nolinkurl{#1}}}
\providecommand{\doeprint}[1]{\href{http://ascl.net/#1}{\nolinkurl{http://ascl.net/#1}}}
\providecommand{\doarXiv}[1]{\href{https://arxiv.org/abs/#1}{\nolinkurl{https://arxiv.org/abs/#1}}}

\end{thebibliography}


\begin{thebibliography}{}
\bibitem[Afferbach et al. (1995)]{Afferbach1995} Afferbach, A., Churchwell, E., \& Werner, M.~W.\ 1995, From Gas to Stars to Dust, Volume 73, pp 111-114
\bibitem[Alloin et al. (1979)]{Alloin1979} Alloin, D., Collin-Souffrin, S., Joly, M., \& Vigroux, L.\ 1979, \aap, 78, 200
\bibitem[Anderson et al.(2011)]{Anderson2011} Anderson, L.~D., Bania, T.~M., Balser, D.~S., Rood, R.~T.\ 2011, \apjs, 194, 32
\bibitem[Anderson et al. (2014)]{Anderson2014} Anderson, L.~D., Bania, T.~M., Balser, D.~S., et al.\ 2014, \apjs, 212, 1
\bibitem[Anderson et al. (2015)]{Anderson2015} Anderson, L.~D., Armentrout, W.~P., Johnstone, B.~M., et al.\ 2015, \apjs, 221, 26
\bibitem[Anderson et al. (2018)]{Anderson2018} Anderson, L.~D., Armentrout, W.~P., Luisi, M., et al.\ 2018, \apjs, 234, 33
\bibitem[Balser et al.(2015)]{Balser2015} Balser, D.~S., Wenger, T.~V., Anderson, L.~D., \& Bania, T.~M.\ 2015, \apj, 806, 199
\bibitem[Bania et al.(2010)]{Bania2010} Bania, T.~M., Anderson, L.~D., Balser, D.~S., Rood, R.~T.\ 2010, \apjl, 718, L106
\bibitem[Brand(1986)]{Brand1986} Brand, J. 1986, PhD thesis, Leiden Univ., The Netherlands
\bibitem[Bresolin(2007)]{Bresolin2007} Bresolin, F.\ 2007, \apj, 656, 186
\bibitem[Chiappini et al. (1997)]{Chiappini1997} Chiappini, C., Matteucci, F., \& Gratton, R.\ 1997, \apj, 477, 765
\bibitem[Cui et al. (2012)]{Cui2012} Cui, X.-Q., Zhao, Y.-H., Chu, Y.-Q., et al.\ 2012, Research in Astronomy and Astrophysics, 12, 1197
\bibitem[Deharveng et al. (2000)]{Deharveng2000} Deharveng, L., Pe{\~n}a, M., Caplan, J., \& Costero, R.\ 2000, \mnras, 311, 329
\bibitem[Esteban et al. (2005)]{Esteban2005} Esteban, C., Garc{\'{\i}}a-Rojas, J., Peimbert, M., et al.\ 2005, \apjl, 618, L95
\bibitem[Esteban et al. (2017)]{Esteban2017} Esteban, C., Fang, X., Garc{\'{\i}}a-Rojas, J., \& Toribio San Cipriano, L.\ 2017, \mnras, 471, 987
\bibitem[Fern{\'a}ndez-Mart{\'{\i}}n et al. (2017)]{FernandezMart2017} Fern{\'a}ndez-Mart{\'{\i}}n, A., P{\'e}rez-Montero, E., V{\'{\i}}lchez, J.~M., \& Mampaso, A.\ 2017, \aap, 597, A84
\bibitem[Fich \& Silkey (1991)]{Fich1991} Fich, M., \& Silkey, M.\ 1991, \apj, 366, 107
\bibitem[Garnett (1992)]{Garnett1992} Garnett, D.~R.\ 1992, \aj, 103, 1330
\bibitem[Hawley (1978)]{Hawley1978} Hawley, S.~A.\ 1978, \apj, 224, 417
\bibitem[Kewley \& Dopita (2002)]{Kewley2002} Kewley, L.~J., \& Dopita, M.~A.\ 2002, \apjs, 142, 35
\bibitem[Kewley \& Ellison (2008)]{Kewley2008} Kewley, L.~J., \& Ellison, S.~L.\ 2008, \apj, 681, 1183-1204
\bibitem[Kniazev et al. (2008)]{Kniazev2008} Kniazev, A.~Y., Pustilnik, S.~A., \& Zucker, D.~B.\ 2008, \mnras, 384, 1045
\bibitem[Korotin et al. (2014)]{Korotin2014} Korotin, S.~A., Andrievsky, S.~M., Luck, R.~E., et al.\ 2014, \mnras, 444, 3301
\bibitem[Lagrois et al. (2012)]{Lagrois2012} Lagrois, D., Joncas, G., \& Drissen, L.\ 2012, \mnras, 420, 2280
\bibitem[L{\'o}pez-S{\'a}nchez et al. (2012)]{Lopez-Sanchez2012} L{\'o}pez-S{\'a}nchez, {\'A}.~R., Dopita, M.~A., Kewley, L.~J., et al.\ 2012, \mnras, 426, 2630
\bibitem[Luo et al. (2015)]{Luo2015} Luo, A.-L., Zhao, Y.-H., Zhao, G., et al.\ 2015, Research in Astronomy and Astrophysics, 15, 1095
\bibitem[Magrini et al. (2003)]{Magrini2003} Magrini, L., Perinotto, M., Corradi, R.~L.~M., \& Mampaso, A.\ 2003, \aap, 400, 511
\bibitem[Marino et al. (2013)]{Marino2013} Marino, R.~A., Rosales-Ortega, F.~F., S{\'a}nchez, S.~F., et al.\ 2013, \aap, 559, A114
\bibitem[Pagel et al.( 1979)]{Pagel1979} Pagel, B.~E.~J., Edmunds, M.~G., Blackwell, D.~E., Chun, M.~S., \& Smith, G.\ 1979, \mnras, 189, 95
\bibitem[P{\'e}rez-Montero (2017)]{PerezMontero2017} P{\'e}rez-Montero, E.\ 2017, \pasp, 129, 043001
\bibitem[Pettini \& Pagel (2004)]{PP2004} Pettini, M., \& Pagel, B.~E.~J.\ 2004, \mnras, 348, L59
\bibitem[Quireza et al. (2006)]{Quireza2006} Quireza, C., Rood, R.~T., Bania, T.~M., Balser, D.~S., \& Maciel, W.~J.\ 2006, \apj, 653, 1226
\bibitem[Riesgo-Tirado \& L{\'o}pez (2002)]{RiesgoTirado2002} Riesgo-Tirado, H., \& L{\'o}pez, J.~A.\ 2002, Revista Mexicana de Astronomia y Astrofisica Conference Series, Ionized Gaseous Nebulae, a Conference to Celebrate the 60th Birthdays of Silvia Torres-Peimbert and Manuel Peimbert, Eds.{Henney}, W.~J. , {Franco}, J., and {Martos}, M.,Mexico City, 12, 174
\bibitem[Rudolph et al. (2006)]{Rudolph2006} Rudolph, A.~L., Fich, M., Bell, G.~R., et al.\ 2006, \apjs, 162, 346
\bibitem[Sabbadin et al. (1977)]{Sabbadin1977} Sabbadin, F., Minello, S., \& Bianchini, A.\ 1977, \aap, 60, 147
\bibitem[Searle(1971)]{Searle1971} Searle, L.\ 1971, \apj, 168, 327
\bibitem[Shaver et al. (1983)]{Shaver1983} Shaver, P.~A., McGee, R.~X., Newton, L.~M., Danks, A.~C., \& Pottasch, S.~R.\ 1983, \mnras, 204, 53
\bibitem[Stasi{\'n}ska (2005)]{Stasinska2005} Stasi{\'n}ska, G.\ 2005, \aap, 434, 507
\bibitem[van Zee et al. (1998)]{van Zee1998} van Zee, L., Salzer, J.~J., Haynes, M.~P., O'Donoghue, A.~A., \& Balonek, T.~J.\ 1998, \aj, 116, 2805
\bibitem[Vanden Berk et al. (2001)]{Vanden Berk2001} Vanden Berk, D.~E., Richards, G.~T., Bauer, A., et al.\ 2001, \aj, 122, 549
\bibitem[V{\'{\i}}lchez \& Esteban (1996)]{Vilchez1996} V{\'{\i}}lchez, J.~M., \& Esteban, C.\ 1996, \mnras, 280, 720
\bibitem[Wang Li-Li et al. (2018)]{Wang L.L.2018} Wang, L.-L., Luo, A.-L., Shen, S.-Y., et al.\ 2018, \mnras, 474, 1873
\bibitem[Yin et al. (2007)]{Yin2007} Yin, S.~Y., Liang, Y.~C., Hammer, F., et al.\ 2007, \aap, 462, 535
\bibitem[Yuan et al. (2015)]{Yuan2015} Yuan, H.-B., Liu, X.-W., Huo, Z.-Y., et al.\ 2015, \mnras, 448, 855
\bibitem[Zhao et al. (2012)]{Zhao2012} Zhao, G., Zhao, Y.-H., Chu, Y.-Q., Jing, Y.-P., \& Deng, L.-C.\ 2012, Research in Astronomy and Astrophysics, 12, 723
\end{thebibliography}
\end{document}